\documentclass[fleqn,usenatbib]{mnras}

\usepackage{newtxtext,newtxmath}

\usepackage[T1]{fontenc}
\usepackage{ae,aecompl}

\usepackage{graphicx}	
\usepackage{amsmath}	

\usepackage{bm, xcolor}
\usepackage[all]{hypcap}
\usepackage{epsfig}
\usepackage{times}

\newcommand\Msun{\,\rmn{M}_{\sun}}

\newcommand\Gyr{\,\rmn{Gyr}}
\newcommand\Myr{\,\rmn{Myr}}

\newcommand\kpc{\,\rmn{kpc}}

\newcommand\MLsun{\,(\rmn{M/L})_{\sun}}

\newcommand\Mvir{M_\rmn{vir}}
\newcommand\FeH{\lbrack \rmn{Fe}/\rmn{H} \rbrack}

\newcommand{\kms}{\,\rmn{km}\,\rmn{s}^{-1}}

\newcommand{\subfind}  {\textsc{subfind}}

\newcommand{\cMpc}     {\,{\rm cMpc}}

\newcommand{\ML}[1]{M/L_\rmn{#1}}

\title[Accreted galaxy properties from GC orbits]{Predicting accreted satellite galaxy masses and accretion redshifts based on globular cluster orbits in the E-MOSAICS simulations}

\author[J. L. Pfeffer et al.]{Joel L. Pfeffer,$^{1}$\thanks{E-mail: \href{j.l.pfeffer@ljmu.ac.uk}{j.l.pfeffer@ljmu.ac.uk}}
Sebastian Trujillo-Gomez,$^{2}$
J. M. Diederik Kruijssen,$^{2}$ \newauthor
Robert A. Crain$^{1}$,
Meghan E. Hughes,$^{1}$
Marta Reina-Campos,$^{2}$
Nate Bastian$^{1}$
\\
$^{1}$Astrophysics Research Institute, Liverpool John Moores University, 146 Brownlow Hill, Liverpool L3 5RF, UK\\
$^{2}$Astronomisches Rechen-Institut, Zentrum f\"{u}r Astronomie der Universit\"{a}t Heidelberg, M\"{o}nchhofstra\ss e 12-14, 69120 Heidelberg, Germany\\
}

\date{Accepted 2020 September 21. Received 2020 July 31; in original form 2020 March 1}

\pubyear{2020}

\begin{document}
\label{firstpage}
\pagerange{\pageref{firstpage}--\pageref{lastpage}}
\maketitle

\begin{abstract}
The ages and metallicities of globular clusters (GCs) are known to be powerful tracers of the properties of their progenitor galaxies, enabling their use in determining the merger histories of galaxies.
However, while useful in separating GCs into individual accretion events, the orbits of GC groups themselves have received less attention as probes of their progenitor galaxy properties.
In this work, we use simulations of galaxies and their GC systems from the E-MOSAICS project to explore how the present-day orbital properties of GCs are related to the properties of their progenitor galaxies. 
We find that the orbits of GCs deposited by accretion events are sensitive to the mass and merger redshift of the satellite galaxy. 
Earlier mergers and larger galaxy masses deposit GCs at smaller median apocentres and lower total orbital energy.
The orbital properties of accreted groups of GCs can therefore be used to infer the properties of their progenitor galaxy, though there exists a degeneracy between galaxy mass and accretion time.
Combining GC orbits with other tracers (GC ages, metallicities) will help to break the galaxy mass/accretion time degeneracy, enabling stronger constraints on the properties of their progenitor galaxy.
In situ GCs generally orbit at lower energies (small apocentres) than accreted GCs, however they exhibit a large tail to high energies and even retrograde orbits (relative to the present-day disc), showing significant overlap with accreted GCs.
Applying the results to Milky Way GCs groups suggests a merger redshift $z \sim 1.5$ for the \textit{Gaia} Sausage/Enceladus and $z>2$ for the `low-energy'/\textit{Kraken} group, adding further evidence that the Milky Way had two significant mergers in its past.
\end{abstract}

\begin{keywords}
galaxies: star clusters: general -- globular clusters: general -- stars: formation -- galaxies: formation -- galaxies: evolution -- methods: numerical
\end{keywords}



\section{Introduction}
\label{sec:intro}

A major objective in astrophysics is to understand the formation and assembly of the Milky Way. 
This has been undertaken from many different angles, combining stellar chemical compositions and ages with their spatial and kinematic properties \citep[e.g.][]{Eggen_Lynden-Bell_and_Sandage_62, Searle_and_Zinn_78, Chiba_and_Beers_00, Carollo_et_al_07}.
These works have since confirmed that the Galaxy formed in a continuous hierarchical process \citep{White_and_Rees_78, Blumenthal_et_al_84} through a combination of \textit{in situ} star formation from accreted and reprocessed gas, and the accretion and merging of satellites which experienced their own chemical evolution \citep[see reviews by][]{Majewski_93, Freeman_and_Bland-Hawthorn_02, Helmi_08}.

Despite ongoing debates about their origin \citep[for recent reviews, see][]{Kruijssen_14, Forbes_et_al_18}, globular clusters (GCs) also have a similarly long history of being used as probes of the galaxy formation and assembly process, particularly in the Milky Way \citep[e.g.][]{Searle_and_Zinn_78, Dinescu_Girard_and_van_Altena_99, Brodie_and_Strader_06}.
The properties of Milky Way GCs, such as their orbits \citep[e.g.][]{Lin_and_Richer_92, Dinescu_Girard_and_van_Altena_99}, metallicities and ages \citep[e.g.][]{Marin-Franch_et_al_09, Forbes_and_Bridges_10, Leaman_VandenBerg_and_Mendel_13, K19b}, horizontal branch morphology \citep[e.g.][]{van_den_Bergh_93, Zinn_93, Mackey_and_Gilmore_04} and chemical abundances \citep[e.g.][]{Brown_Wallerstein_and_Zucker_97, Pritzl_Venn_and_Irwin_05, Horta_et_al_20}, have been used to distinguish their origin (accretion or in situ formation) and derive the properties of their progenitor galaxies which have long since been accreted.
The orbital properties of GCs, in particular, hold great promise in separating the GCs of individual accretion events if their orbital properties remain clustered (e.g. in integrals of motion).

The release of proper motions from the \textit{Gaia} mission \citep{Gaia_DR1_16,Gaia_DR2_18}, in particular the second data release, has seen a major advance in deriving the orbits of field stars and GCs and generated a large number of works on the assembly history of the Milky Way and the origin of its GC population \citep[e.g.][]{Belokurov_et_al_18, Belokurov_et_al_20, Deason_et_al_18, Deason_Belokurov_and_Sanders_19, Haywood_et_al_18, Helmi_et_al_18, Myeong_et_al_18a, Myeong_et_al_18b, Myeong_et_al_18c, Myeong_et_al_18d, Myeong_et_al_19, Di_Matteo_et_al_19, Iorio_and_Belokurov_19, Koppelman_et_al_19a, Koppelman_et_al_19b, Mackereth_et_al_19, Massari_et_al_19, Necib_et_al_19, Necib_et_al_20, Vasiliev_19}.
A major outcome of these works is that the outer stellar halo ($>10\kpc$) of the Milky Way appears to be dominated by the debris from a single merged satellite, the \textit{Gaia} Sausage/Enceladus (G-E), which was accreted $\sim 9$-$10\Gyr$ ago \citep{Belokurov_et_al_18, Haywood_et_al_18, Helmi_et_al_18, Myeong_et_al_18d, Myeong_et_al_19, Bignone_Helmi_and_Tissera_19, Conroy_et_al_19b, Mackereth_et_al_19, Kruijssen_et_al_20}.
Along with G-E, the analysis of \textit{Gaia} DR2 data has resulted in the discovery of a number of less massive substructures and the characterisation of their progenitor galaxies (e.g. Sequoia, \citealt{Myeong_et_al_19}; Thamnos, \citealt{Koppelman_et_al_19b}) and further constrained the progenitor properties of already known substructures \citep[e.g. Helmi streams,][]{Helmi_et_al_99, Koppelman_et_al_19a}.

\citet{Massari_et_al_19} also found evidence for a population of GCs in the Milky Way at low energies (the L-E group) which do not appear to have formed in situ in the Galaxy (the GCs have ages and metallicities consistent with the `accreted' or `satellite' branch\footnote{The age-metallicity relation of the Milky Way GCs appears to be bifurcated, with the young, metal-poor branch thought to originate from the accretion of satellite galaxies and their GCs \citep{Marin-Franch_et_al_09, Forbes_and_Bridges_10, Leaman_VandenBerg_and_Mendel_13, K19b}.}), and are not connected to previously known merger events.
This GC population is plausibly consistent with the proposed \textit{Kraken} \citep{K19b} accretion event, which was predicted based on number of GCs in the `satellite' branch of the Milky Way GC age-metallicity distribution \citep[see also][]{Forbes_20,Kruijssen_et_al_20}.

In this work, we use the hydrodynamical, cosmological simulations of galaxy formation including GC formation and evolution from the MOdelling Star cluster population Assembly In Cosmological Simulations within EAGLE project \citep[E-MOSAICS][]{P18,K19a} to investigate the orbits of accreted and in situ GCs.
The E-MOSAICS simulations have previously been used to investigate the origin and evolution of GCs \citep{P18,Reina-Campos_et_al_18,Reina-Campos_et_al_19,Usher_et_al_18,Hughes_et_al_20, Keller_et_al_20}, the use of GCs to trace the formation and assembly of galaxies \citep{K19a,K19b,Hughes_et_al_19,Trujillo-Gomez_et_al_20} and the properties of young clusters at high and low redshifts \citep{P19a,P19b}.
This work explores how the orbital properties of accreted GCs can be related to their progenitor galaxy properties (mass and accretion time). 
In this paper, we aim to test which types of galaxy accretion events could place GCs on orbits similar to the G-E and L-E groups, and test whether the latter is consistent with the proposed Kraken accretion event.
Additionally, we compare the orbital properties of GCs formed in situ within the main progenitor galaxies and investigate possible overlap between in situ and accreted clusters.
In a companion paper \citep{Kruijssen_et_al_20}, we extend this analysis, combining the orbital properties of GC sub-groups with their ages and metallicities to predict the properties of their progenitor galaxies.

The paper is organised as follows. 
Section \ref{sec:methods} describes the E-MOSAICS simulations and the analysis of their results.
Section \ref{sec:results} presents the main results of this work, comparing the orbital properties of the simulated GC populations with those of the Milky Way GCs.
Finally, we discuss and summarize our findings in Section \ref{sec:discussion}.

\section{Methods}
\label{sec:methods}

\subsection{Simulations}
\label{sec:simulations}

The E-MOSAICS project is a suite of cosmological, hydrodynamical simulations of galaxy formation in the $\Lambda$ cold dark matter cosmogony, which includes a subgrid model for star cluster formation and evolution \citep[we refer the reader to these works for a full description of the models and simulations]{P18,K19a}.
E-MOSAICS couples MOSAICS star cluster model \citep{Kruijssen_et_al_11,P18} to the Evolution and Assembly of GaLaxies and their Environments (EAGLE) model for galaxy formation \citep{S15,C15}.
The EAGLE model reproduces a wide range of galaxy properties, including the redshift evolution of galaxy stellar masses, specific star formation rates and sizes \citep{Furlong_et_al_15, Furlong_et_al_17}, galaxy luminosities and colours \citep{Trayford_et_al_15_short}, their cold gas properties \citep{Lagos_et_al_15_short, Lagos_et_al_16, Bahe_et_al_16, Marasco_et_al_16, Crain_et_al_17} and the chemical abundance patterns observed in the Milky Way \citep{Mackereth_et_al_18, Hughes_et_al_20}.

The MOSAICS model treats star cluster formation and evolution in a subgrid fashion, such that clusters are `attached' to stellar particles. 
Star clusters are spawned at the time of formation of a stellar particle and adopt the basic properties of their host particle (positions, velocities, ages, abundances).
The formation (numbers, masses) and evolution (mass loss) of clusters is governed by local properties within the simulations (gas density and pressure, tidal field).
MOSAICS adopts a cluster formation model \citep{Kruijssen_12, Reina-Campos_and_Kruijssen_17} that has been shown to reproduce the observed properties of young star cluster populations in nearby galaxies \citep{P19b}. 
Following their formation, star clusters may then lose mass due to stellar evolution (according to the EAGLE model), tidal shocks, two-body relaxation or may be completely removed due to dynamical friction in the host galaxy (the latter is treated in post-processing, meaning particle orbits are not modified).

In this work, we analyse the volume-limited set of 25 simulations of Milky Way-mass haloes ($\Mvir \approx 10^{12} \Msun$) from the E-MOSAICS project \citep{P18,K19a}.
The haloes were drawn from the high-resolution $25 \cMpc$ volume EAGLE simulation \citep[Recal-L025N0752;][]{S15} and resimulated in a zoom-in fashion with the same parameters as the parent volume (a \citealt{Planck_2014_paperI} cosmology, the `recalibrated' EAGLE model and initial baryonic particle masses of $\approx2.25 \times 10^5 \Msun$).
In total, 29 snapshots are produced between $z=20$ and $z=0$ for each simulation.
Bound galaxies (subhaloes) are identified at each snapshot using the \subfind\ algorithm \citep{Springel_et_al_01, Dolag_et_al_09} and merger trees for the subhaloes are created using the method described in \citet{P18}.

\subsection{Analysis}
\label{sec:analysis}

We limit GCs in the simulations to star clusters with masses $>5 \times 10^{4} \Msun$ at $z=0$ and metallicities $-2.5 < \FeH < -0.5$.
This selection is similar to the properties of Milky Way star clusters for which orbital properties \citep[e.g.][]{Myeong_et_al_18d, Baumgardt_et_al_19, Massari_et_al_19} and ages have been determined \citep{Forbes_and_Bridges_10, Dotter_et_al_10, Dotter_Sarajedini_and_Anderson_11, VandenBerg_et_al_13}.
The upper metallicity limit also mitigates the over-survival of metal-rich clusters in the simulations \citep[see][for discussion]{K19a}.

We calculate peri- ($r_\rmn{peri}$) and apocentres ($r_\rmn{apo}$) for the orbits of these GCs at $z=0$ and at the snapshot immediately after the stellar particle was formed (the `initial' value) following the method described in \citet[see section 2.4]{Mackereth_et_al_19}\footnote{This method assumes the potential is approximately spherically symmetric, which is not assumed for the calculation of $E$ and $L_z$ below. However, this is a reasonable assumption in the region where the disc/bulge does not dominate the potential, since the effect of dissipation in baryonic simulations makes dark matter haloes (which dominate the potential in the galaxies) significantly more spherical than in dark matter only simulations \citep{Dubinski_94, Kazantzidis_et_al_04, Springel_White_and_Hernquist_04}.}.
Clusters with unbound orbits at a given snapshot are disregarded, since their orbital parameters cannot be determined.
Eccentricities of the orbits are calculated as $e = (r_\rmn{apo}-r_\rmn{peri})/(r_\rmn{apo}+r_\rmn{peri})$.

The angular momentum and total energy of the GCs are calculated at $z=0$ as follows. For the $z$-component of the angular momentum, $L_z$, we first rotate each galaxy to align the total angular momentum of all the bound stars (as calculated by \subfind) with the $z$-axis, assuming that it corresponds to the symmetry axis of the potential. We verified that this is typically the case for galaxies with a clear disc-like morphology. The value of $L_z$ for each GC is then obtained by projecting its total angular momentum vector onto the $z$-axis. The total energy is obtained by adding the kinetic energy to the gravitational potential energy at the position of each object in the $z=0$ snapshot. The gravitational potential is calculated by doing a direct sum over the contributions from all the particles in the simulation box. Both the angular momentum and the total energy are expressed per unit mass.

In order to trace accretion events of GCs onto the main galaxy during the simulations, stellar particles that are bound to the main galaxy at $z=0$ must first be associated to a `parent' galaxy or branch in the merger tree.
For particles bound to subhaloes in the same branch both prior to and after star formation occurs, association is trivial (and stars/GCs are clearly formed in situ or are accreted).
When a particle changes galaxy branch between the snapshots prior to and after star formation (i.e. during a galaxy merger), associating it to a parent galaxy is less straightforward.
The maximum time between snapshots for the simulations is $1.35 \Gyr$ (at $z \approx 0$), much larger than the typical dynamical timescale for a particle in the central subhalo ($\sim 100 \Myr$ at a galactocentric radius of $10$-$20 \kpc$ in a Milky Way-mass halo, depending on redshift and galaxy mass).
Therefore, a gas particle may be accreted from a satellite and become dynamically associated with the central subhalo on a timescale shorter than that between snapshots.
For this reason, we define the parent subhalo as the subhalo the particle was bound to at the snapshot $<100 \Myr$ prior to the particle becoming a star, if one exists, and otherwise at the snapshot immediately after star formation.\footnote{Note that, depending on the timing of the galaxy merger relative to the snapshots, the method could potentially under-associate particles to an accreting galaxy. This method would be improved simply by taking a higher frequency of snapshots.}
Where multiple snapshots fall within $100 \Myr$ prior to star formation (which is possible at $z>8$), we define the parent subhalo as the subhalo with the lowest branch mass (generally the earliest accreted branch).
To define the accretion event during which a GC was accreted into the main galaxy, we trace the merger tree from the parent subhalo to the main branch of the merger tree for the central galaxy at $z=0$.

In situ GCs are also defined based on their parent subhalo, with the addition of two criteria: a galactocentric radius selection and the requirement that the particle was bound to the main galaxy branch prior to becoming a star particle.
This allows us to define a sample of \textit{clearly} in situ GCs and excludes those with an ambiguous origin.
Following \citet{Sanderson_et_al_18}, we define in situ star particles as those located within $30 \kpc$ of the main galaxy at the snapshot immediately after star formation.
This selection therefore excludes (e.g.) stellar particles that may have formed in the tidal tails of accreting galaxies, but which are not bound to the incoming satellite at the time of formation as determined by the \subfind\ algorithm, and would otherwise be classified as in situ formation.
The combination of the radius selection and being bound to the central galaxy prior to star formation excludes $\approx 12$ per cent of GCs with an ambiguous origin (the majority of which, $7$ per cent, do not pass the radius cut), which would be classed as in situ formation by the merger tree criteria alone.

To compare the in situ GCs from the simulations against the Milky Way GCs (Section \ref{sec:insitu}), we select galaxies from the 25 zoom-in simulations with $z=0$ properties most similar to the Milky Way, excluding spheroidal galaxies and those with late major mergers.
We first select disc-dominated galaxies with a disc-to-total stellar mass $D/T > 0.45$ \citep[equivalent to the fraction of kinetic energy invested in ordered corotation $\kappa_\rmn{co} = 0.4$,][]{Correa_et_al_17, Thob_et_al_19}.
Disc stars are selected following \citet{Abadi_et_al_03b} and \citet{Sales_et_al_12} and require an orbit circularity parameter $J_\rmn{z} / J_\rmn{circ} > 0.5$ (i.e. the ratio of the specific angular momentum perpendicular to the disc to that of a circular orbit with the same energy).
We also remove galaxies that have on-going (at $z=0$) or late major mergers ($z<0.8$) with a stellar mass ratio $M_2/M_1 > 1/4$ (where $M_2 < M_1$).
These criteria leave us with 14 galaxies from the sample for which we compare in situ GCs. For the accreted GCs we use all 25 Milky Way-mass galaxies, since we do not expect the GC orbits to be strongly affected by the present day galaxy morphology.

\section{Results} 
\label{sec:results}

\subsection{Milky Way GCs}
\label{sec:MW}

\begin{figure}
  \includegraphics[width=84mm]{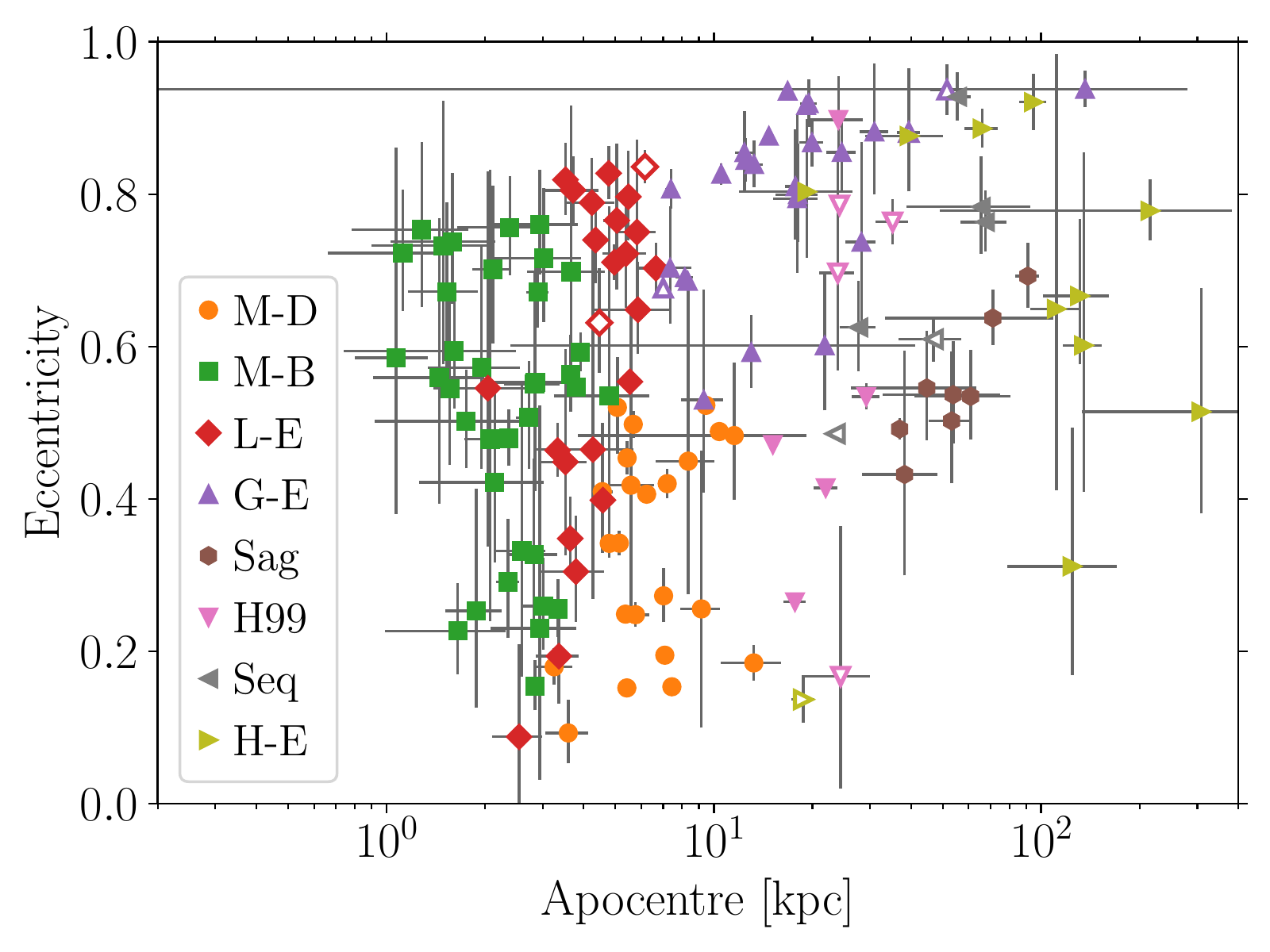}
  \includegraphics[width=84mm]{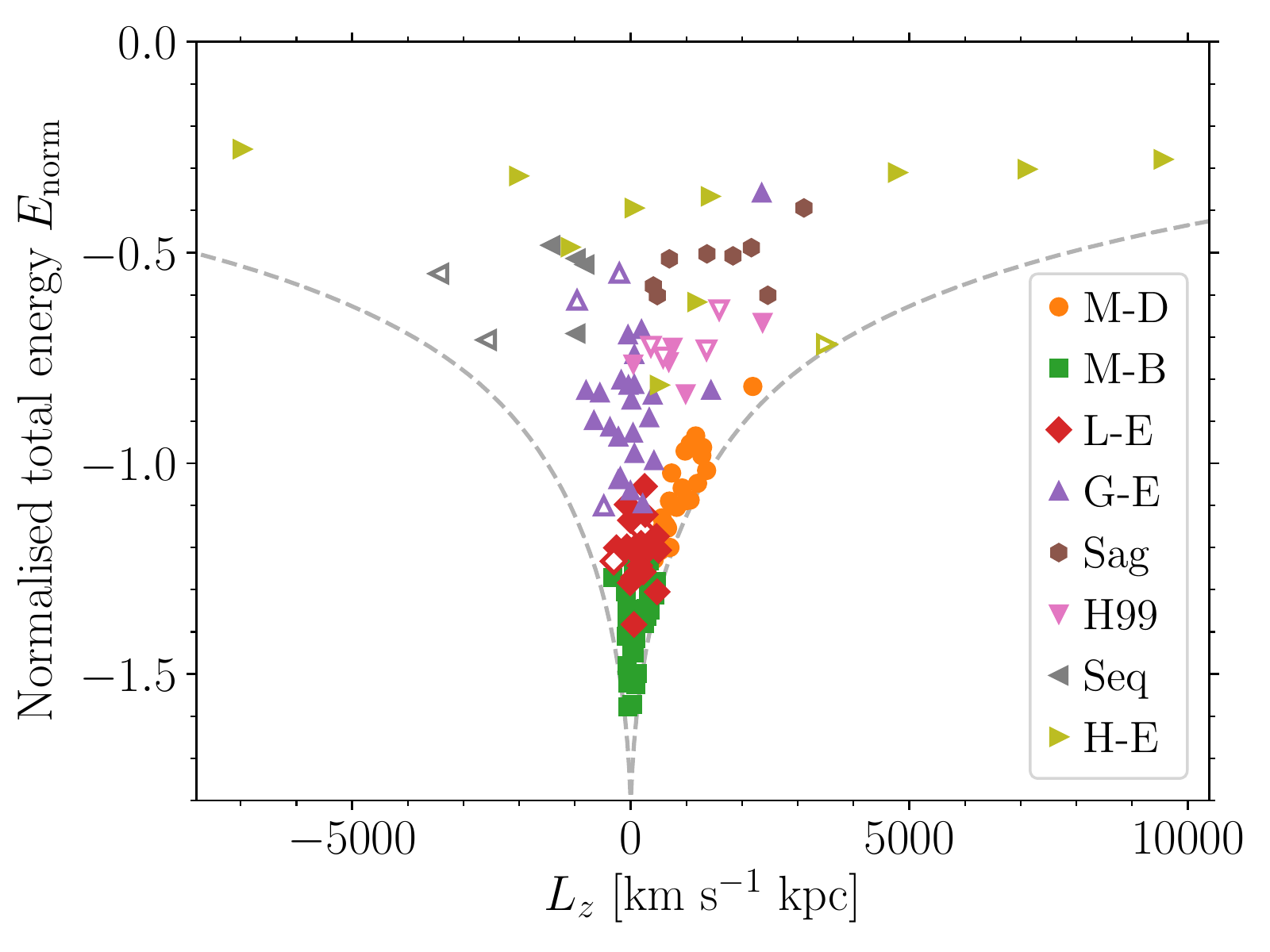}
  \caption{Orbital properties of Milky Way GCs. The upper panel shows apocentres and eccentricities \citep[from][]{Baumgardt_et_al_19} while the lower panel shows the normalised total energy (normalised at $L_z = 1500 \kms \kpc$) and the $z$-component of the angular momentum. GC groupings were taken from \citet[M-D: Main-disc, M-B: Main-Bulge, L-E: low energy/Kraken, G-E: \textit{Gaia} Sausage/Enceladus, Sag: Sagittarius dSph, H99: \citet{Helmi_et_al_99} streams, Seq: Sequoia, H-E: high energy; with some minor changes, see text]{Massari_et_al_19}. Open symbols show clusters with uncertain group designations.}
  \label{fig:MW_GCs}
\end{figure}

For reference and comparison with the simulations, in Fig. \ref{fig:MW_GCs} we show the  orbital properties of Milky Way GCs. 
In the upper panel we show their apocentres and eccentricities, while in the lower panel we show their total energy (normalized to $E_\mathrm{norm}=-1$ at $L_z = 1500 \kms \kpc$ for comparison with the simulations, see Section \ref{sec:E-Lz}) and $z$-component of the angular momentum.
The apocentres and eccentricities were taken from \citet{Baumgardt_et_al_19}. The angular momentum was calculated using the velocities and distances from \citet{Baumgardt_et_al_19}, and the energies were obtained assuming the \citet{McMillan_17} potential model of the Milky Way\footnote{\citet{Baumgardt_et_al_19} assume the \citet{Irrgang_et_al_13} potential model to integrate the orbits. However, they note that there was little difference in the results when assuming the \citet{McMillan_17} model.}.

We take the groupings into possible progenitors from \citet[M-D: Main-disc, M-B: Main-Bulge, L-E: low energy/Kraken, G-E: \textit{Gaia} Sausage/Enceladus, Sag: Sagittarius dSph, H99: \citet{Helmi_et_al_99} streams, Seq: Sequoia, H-E: high energy]{Massari_et_al_19}, with a few modifications. 
Based on the GC ages and metallicities in the compilation of \citet{K19b}, we updated the group associations of E3 to M-D, NGC 6441 to M-B and labelled Palomar 1 (H-E) and NGC 6121 (L-E) as uncertain (Palomar 1 has an age/metallicity consistent with young satellite GCs but an orbit consistent with disc GCs, while NGC 6121 has an age placing it intermediate between the accreted and in situ branches in age-metallicity space). 
See \citet{Kruijssen_et_al_20} for further discussion about the memberships of these GCs.
We note that the division of GCs into accretion groups is always somewhat uncertain, in particular given possible overlap of the groups in orbital space, and subject to change given new information. We touch on this point in Section \ref{sec:indiv_gals}, where we present the orbital properties of GCs accreted by individual galaxies.
In general we will compare the median properties of possible GC accretion groups, since the median properties are relatively robust to the uncertainties of group classification \citep[see also][where we explicitly take this into account]{Kruijssen_et_al_20}.

Though the L-E and G-E GC groups have similar age-metallicity relations \citep{Massari_et_al_19}, the groups have very different orbital properties, suggesting different origins.
The L-E group has a median apocentre of $4.4 \kpc$ and median eccentricity of $0.68$, while the G-E group has a median apocentre of $18.0 \kpc$ and median eccentricity of $0.81$.
The groups also occupy a very different range in eccentricities: $\approx$0.1-0.85 for the L-E group, compared with $\approx$0.5-0.95 for the G-E group.
The GCs from lower mass progenitors are generally found at larger apocentres (median of $24.0 \kpc$ for the H99 streams, $37.3 \kpc$ for Sequoia and $53.6 \kpc$ for Sagittarius).

\subsection{Orbital trends with mass and merger redshift}
\label{sec:trends}

\subsubsection{Apocentre and eccentricity}
\label{sec:apo-ecc}

\begin{figure*}
  \includegraphics[width=\textwidth]{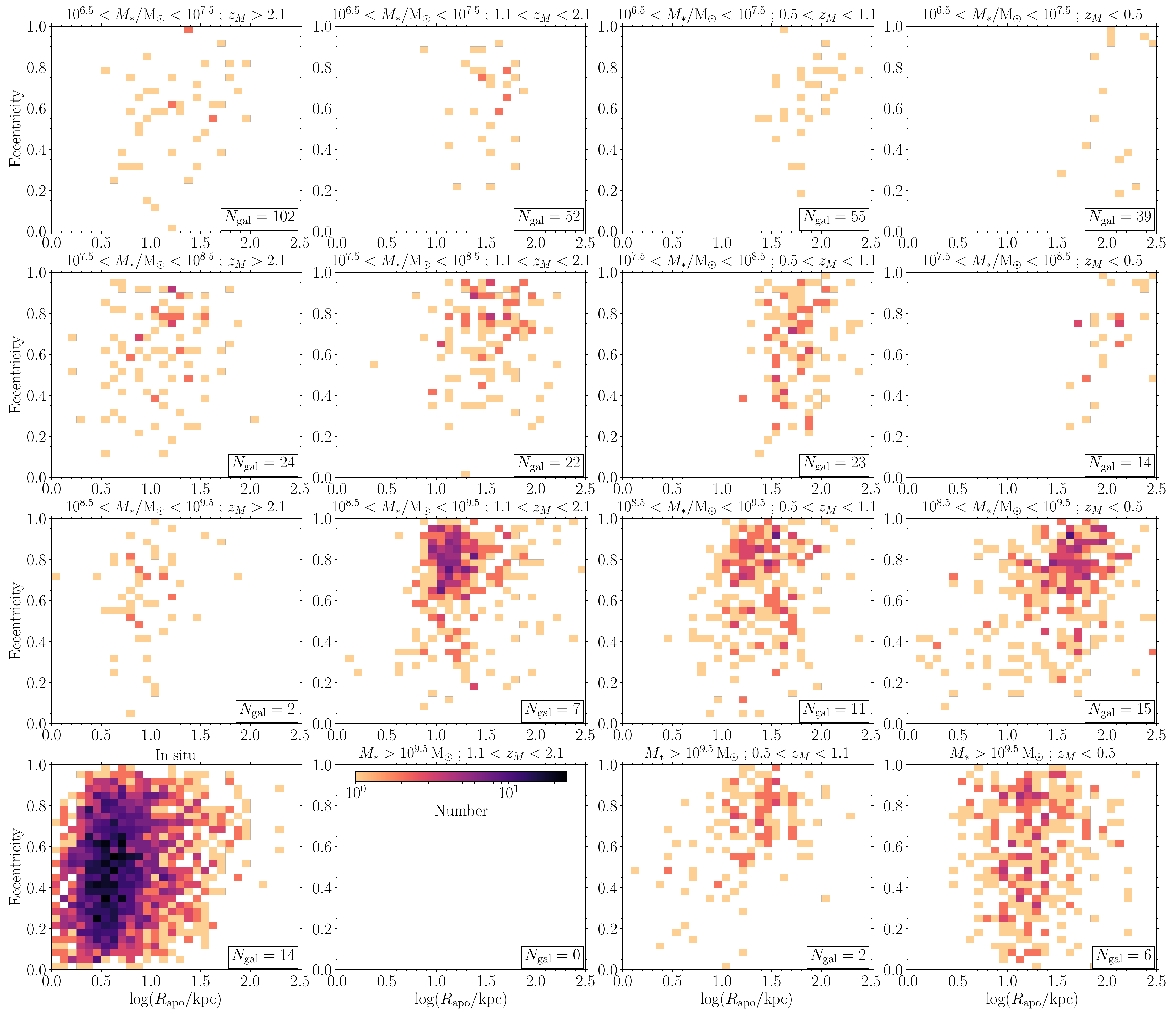}
  \caption{Eccentricities and apocentres of accreted GCs at $z=0$, stacked by satellite galaxy stellar mass (galaxy mass increases from top to bottom) and merger redshift (early to late mergers from left to right) in the 25 Milky Way-mass galaxies from the E-MOSAICS simulations. The bottom left panel shows the eccentricities and apocentres for in situ GCs in disc-dominated galaxies at $z=0$. The colour scale of the histograms is logarithmic, normalized to the maximum in the in situ panel. The inset $N_\rmn{gal}$ shows the number of galaxies which satisfy the $M_\ast$ and $z_M$ selection criteria (for the low mass galaxies, $10^{6.5} < M_\ast / \Msun < 10^{7.5}$, not all galaxies contribute GCs).}
  \label{fig:apo-ecc}
\end{figure*}

\begin{table}
 \caption{Median apocentres (kpc) for GCs of all accreted galaxies in bins of satellite stellar mass and accretion redshift (Fig. \ref{fig:apo-ecc}).}
 \label{tab:apo}
 \begin{tabular}{ccccc}
  \hline
       & \multicolumn{4}{c}{$z_M$} \\
  $M_\ast$ ($\Msun$)    & $>2.1$ & 2.1-1.1 & 1.1-0.5 & $<0.5$ \\
  \hline                                             
  $10^{6.5}$-$10^{7.5}$ & 17.6   & 35.9    & 66.6    & 117.4  \\
  $10^{7.5}$-$10^{8.5}$ & 13.7   & 32.6    & 54.0    & 100.7  \\
  $10^{8.5}$-$10^{9.5}$ &  8.2   & 15.0    & 24.3    &  41.4  \\
           $> 10^{9.5}$ &        &         & 21.9    &  16.5  \\
  \hline
 \end{tabular}
\end{table}

\begin{table}
 \caption{Median eccentricities for GCs of all accreted galaxies in bins of satellite stellar mass and accretion redshift (Fig. \ref{fig:apo-ecc}).}
 \label{tab:ecc}
 \begin{tabular}{ccccc}
  \hline
       & \multicolumn{4}{c}{$z_M$} \\
  $M_\ast$ ($\Msun$)    & $>2.1$ & 2.1-1.1 & 1.1-0.5 & $<0.5$ \\
  \hline                                             
  $10^{6.5}$-$10^{7.5}$ & 0.59   & 0.70    & 0.70    & 0.62   \\
  $10^{7.5}$-$10^{8.5}$ & 0.66   & 0.74    & 0.68    & 0.74   \\
  $10^{8.5}$-$10^{9.5}$ & 0.66   & 0.76    & 0.72    & 0.73   \\
           $> 10^{9.5}$ &        &         & 0.73    & 0.62   \\
  \hline
 \end{tabular}
\end{table}

In Fig. \ref{fig:apo-ecc}, we compare the apocentres and eccentricities of accreted GCs at $z=0$ from the 25 Milky Way-mass galaxies (with the exception of the bottom left panel, which shows in situ GCs and is discussed in Section \ref{sec:insitu}). 
The GCs of accreted galaxies are divided into panels by the satellite galaxy stellar mass at accretion (with increasing mass from upper to lower rows) and the merger redshift (with decreasing merger redshift from left to right columns).
We give the median apocentres and eccentricities for each panel in Tables \ref{tab:apo} and \ref{tab:ecc}, respectively.
Note that, because the effect of dynamical friction on GC orbits is not included, apocentres may be larger than they should be realistically, particularly for GCs that orbit at small galactocentric radii.
For a GC with mass $10^5 \Msun$ on a circular orbit at $2 \kpc$ in MW04, the typical dynamical friction timescale is $t_\mathrm{df} \approx100 \Gyr$. 
Therefore, for a typical GC, this correction is not relevant and only applicable for massive GCs ($\gtrsim10^6 \Msun$, for which $t_\mathrm{df} \lesssim 10 \Gyr$).
However, what cannot be captured in the model is the effect of shrinking orbits within the host satellite prior to merging, which could potentially result in GCs being deposited at smaller apocentres through later tidal stripping.
Another factor which may affect the resulting orbits is the size of the galaxies.
Low mass galaxies ($M_\ast < 10^9 \Msun$) in the EAGLE model are slightly too extended compared to observed galaxies \citep{S15, Furlong_et_al_17}, though the comparison is improved for the `Recalibrated' EAGLE model (used in this work) relative to the lower resolution `Reference' model.
Galaxies which are too extended may suffer from premature tidal disruption relative to more compact galaxies, thus resulting in larger apocentres due to the decreased efficiency of dynamical friction.

For the accreted GCs, at fixed galaxy mass, there is a strong trend of apocentre with merger redshift, with earlier mergers having smaller apocentres.
At fixed merger redshift, apocentres also become smaller with increasing satellite galaxy mass, i.e. mergers with more massive accretors deposit their clusters and stars at smaller apocentres.
Both trends persist across all galaxy mass and merger redshift ranges, respectively (Table \ref{tab:apo}).
The trend with redshift appears reversed for the most massive galaxies ($M_\ast > 10^{9.5} \Msun$) which is likely due to poor galaxy number statistics in the $0.5 - 1.1$ redshift range (two galaxies).
With such small numbers sampling of the galaxy mass function becomes important due to the correlation of galaxy mass and median apocentre after accretion.
Comparing just the most massive accreted galaxy in each redshift range ($M_\ast \approx 10^{10} \Msun$ for both) we find the trend still holds; the galaxy accreted at $z \approx 1$ has a median GC apocentre of $9\kpc$, while the galaxy accreted at $z \approx 0$ has a median apocentre of $19\kpc$.
Other factors, such as the initial orbital conditions of the mergers, may also affect the correlations when sampling of galaxies is poor.

We find no trends for $M_\ast$ or $z_M$ with the median eccentricity of the GC orbits (Table \ref{tab:ecc}; the interquartile range of eccentricity for individual galaxies does however correlate with galaxy mass, which we discuss further in Section \ref{sec:indiv_gals}).
The uncertainties of the medians range from $0.015$ to $0.1$ and thus most bins in redshift and galaxy mass are consistent with the median for all accreted GCs ($0.71 \pm 0.01$).
The median eccentricity for accreted GCs in the simulations ($0.71$) is in extremely good agreement with the median for accreted Milky Way GCs ($0.70 \pm 0.03$ for those not associated with M-D or M-B; Fig. \ref{fig:MW_GCs}).

The cause of the decrease of GC apocentres with increasing progenitor satellite mass and merger redshift is the competition of dynamical friction between the central and satellite galaxies (which occurs self-consistently in the simulations) and tidal stripping of the satellite galaxy.
Dynamical friction occurs most efficiently as the merging galaxies approach a 1:1 mass ratio \citep{Chandrasekhar_43, Binney_and_Tremaine_08}. 
Therefore, for a given central galaxy mass, higher mass galaxies will sink to the centre of the central galaxy on a shorter timescale.
Conversely, lower mass galaxies would also sink to the centre of the central galaxy given enough time, but are tidally stripped, and eventually completely disrupted, on a timescale much faster than the timescale for dynamical friction due to their lower binding energies.
At later times the central galaxy (or dark matter halo) is more massive and therefore, at a given satellite galaxy mass and size, tidal stripping of both field stars and GCs occurs at larger galactocentric radii at lower redshifts.
At the same time, dynamical friction is also less effective at later times as the mass ratio decreases.

Comparing the results of Fig. \ref{fig:apo-ecc} with Fig. \ref{fig:MW_GCs}, we can estimate approximate merger times for the progenitors of the different GC groups in the Milky Way.
The L-E group has a median apocentre of $4.4 \kpc$ (range of $2.7$-$8.7 \kpc$), suggesting it was most likely accreted into the Milky Way at early times ($z>2$).
Late accretion ($z < 1$) is disfavoured because GCs could only be deposited in galaxies at such small apocentres ($<10 \kpc$) during major mergers, for which there is no evidence in the Milky Way \citep{Wyse_01, Hammer_et_al_07, Stewart_et_al_08, K19b}.
The L-E group also favours galaxy masses $M_\ast > 10^{7.5} \Msun$, since clusters accreted from lower mass galaxies at $z>2$ typically have apocentres $>10 \kpc$ (Table \ref{tab:apo}) and cover a much wider range in apocentres (at $z>2$ the interquartile range of the apocentres decreases from $27 \kpc$ in the lowest mass bin to $6 \kpc$ in the highest mass bin).

The \textit{Gaia} Sausage/Enceladus accretion event has been suggested to have a stellar mass $\sim 10^9 \Msun$ and accreted $\sim 9$-$10\Gyr$ ago or later \citep{Belokurov_et_al_18, Haywood_et_al_18, Helmi_et_al_18, Myeong_et_al_18d, Myeong_et_al_19, Bignone_Helmi_and_Tissera_19, Conroy_et_al_19b, Mackereth_et_al_19}, though in \citet{Kruijssen_et_al_20} we find a mass $M_\ast \approx 10^{8.4} \Msun$.
The median apocentre of G-E GCs ($18 \kpc$) does not itself place a constraint on the galaxy mass, since such apocentres are achievable (Table \ref{tab:apo}) through both early, low mass mergers ($z_M > 2$, $M_\ast < 10^{7.5} \Msun$) and late, major mergers ($z_M < 0.5$, $M_\ast > 10^{9.5} \Msun$).
Further information is therefore required to derive a merger time for G-E due to this mass-merger redshift degeneracy.
However, a mass range $10^{7.5} < M/\Msun < 10^{9.5}$ suggests a merger between redshifts $\approx1$-$2$ ($8$-$10.5$ Gyr ago).

The Sagittarius GCs have a median apocentre of $53.6 \kpc$ (Fig. \ref{fig:MW_GCs}), while the Sagittarius dwarf galaxy has a total progenitor luminosity of $M_\rmn{V} \sim -15.1$ to $-15.5$ \citep{Niederste-Ostholt_et_al_10}, or a stellar mass $\sim 2$-$3 \times 10^8 \Msun$ for a stellar mass-to-light ratio of $\ML{V} = 2 \MLsun$\footnote{Assuming an age of $8 \Gyr$ and $\FeH = -0.5$ \citep{Bellazzini_et_al_06} for a simple stellar population with the Flexible Stellar Population Synthesis model \citep*{Conroy_Gunn_and_White_09, Conroy_and_Gunn_10}.}.
From the results of Fig. \ref{fig:apo-ecc} and Table \ref{tab:apo}, these parameters imply a merger redshift $z_M < 1$, consistent with the status of Sagittarius currently undergoing tidal disruption \citep{Ibata_Gilmore_and_Irwin_94, Velazquez_and_White_95, Ibata_et_al_97}.

The H99 stream and Sequoia GC groups both have median apocentres around $30 \kpc$ ($24.0$ and $37.3 \kpc$, respectively).
However, the derived stellar masses differ by an order of magnitude. \citet{Koppelman_et_al_19a} find a stellar mass for the H99 stream progenitor of $M_\ast \sim 10^8 \Msun$, while \citet{Myeong_et_al_19} find a stellar mass for Sequoia of $M_\ast \sim 5$-$70 \times 10^6 \Msun$.
This implies a merger redshift $z_M > 1$ for both galaxies: $z_M \sim 2$ for the H99 stream and $z_M \sim 1.5$ for Sequoia (a lower mass for the H99 stream progenitor, e.g. $10^7 \Msun$, would not significantly change this result), consistent with the youngest ages of probable H99 stream and Sequoia GCs \citep[$\gtrsim 11 \Gyr$,][]{Koppelman_et_al_19a, Massari_et_al_19, Myeong_et_al_19}.

\subsubsection{Energy and angular momentum}
\label{sec:E-Lz}

\begin{figure*}
  \includegraphics[width=\textwidth]{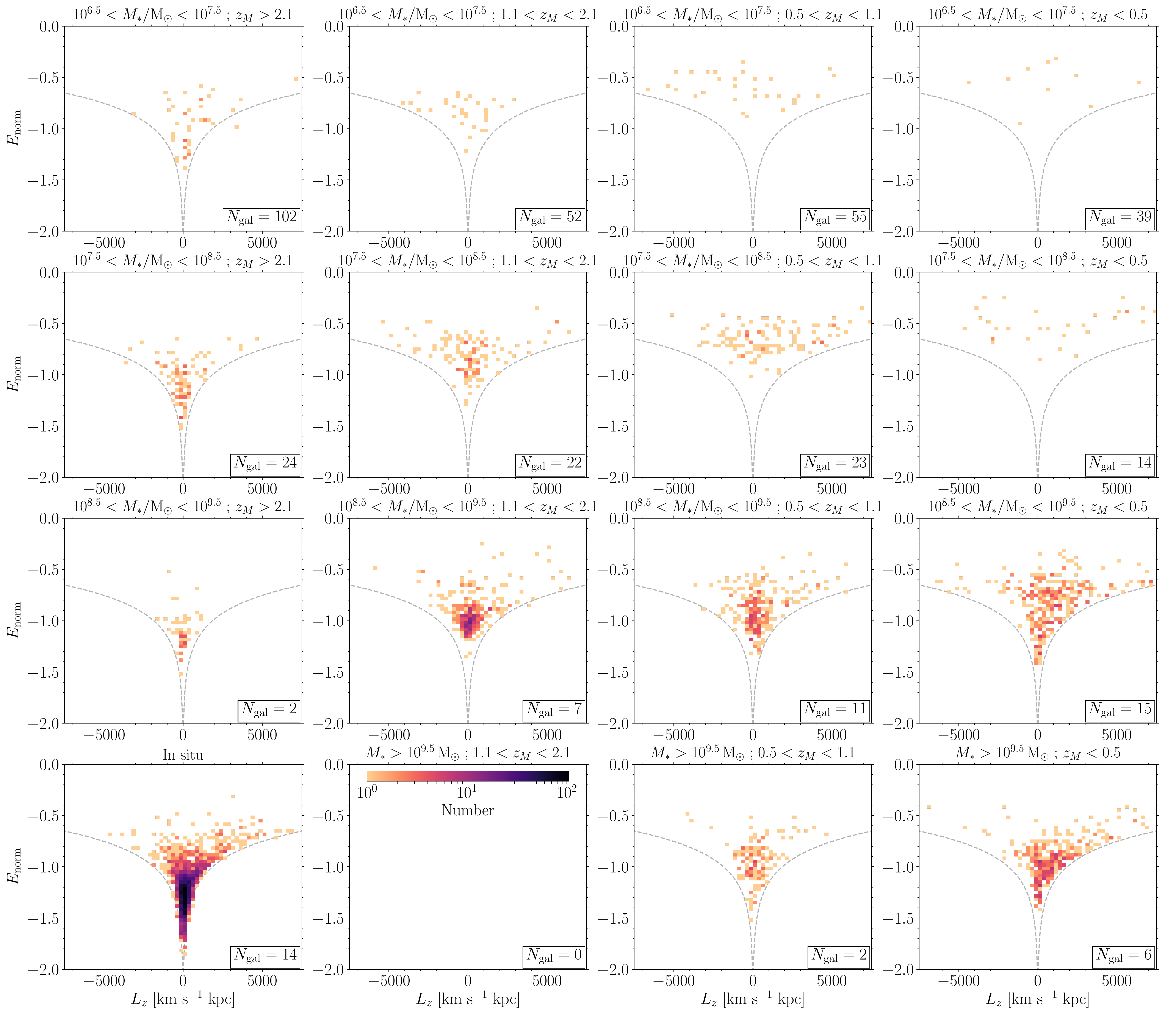}
  \caption{Normalised total energy, $E_\rmn{norm}$ (normalised to $E_\rmn{norm}=-1$ at $L_z = 1500 \kms \kpc$), and the $z$-component of the angular momentum, $L_z$, of accreted GCs at $z=0$. Panel layout is as in Fig. \ref{fig:apo-ecc}, with galaxies stacked by satellite galaxy stellar mass (galaxy mass increases from top to bottom) and merger redshift (early to late mergers from left to right). The bottom left panel shows $E_\rmn{norm}$ and $L_z$ for in situ GCs in disc-dominated galaxies at $z=0$. The colour scale of the histograms is logarithmic, normalized to the maximum in the in situ panel. The dashed lines approximately indicate the circular orbit curve for reference (this is not a fit, since the circular orbit curve differs from galaxy to galaxy).}
  \label{fig:E-Lz}
\end{figure*}

In Fig. \ref{fig:E-Lz}, we show the normalised total energy ($E_\rmn{norm}$) and $z$-component of the angular momentum ($L_z$) for GCs at $z=0$ for the same panels in Fig. \ref{fig:apo-ecc}.
Since the circular orbit curve differs for each galaxy depending on the mass profile, we normalise the energy for the GCs of each galaxy by the absolute value of the energy of a circular orbit with $L_z = 1500 \kms \kpc$\footnote{In practice, we use the minimum $E$ for all star particles with $L_z > 1500 \kms \kpc$.} such that the circular orbit curves are approximately aligned for all galaxies.
Given that the total energy and apocentre for an orbit in a galaxy are related, the typical energy for accreted clusters follows the same trend with galaxy mass and accretion redshift as for the typical apocentre. At fixed satellite mass, GCs from earlier mergers are more tightly bound (lower total energy) than those accreted later; while at fixed merger redshift, GCs from higher mass satellite galaxies are more tightly bound than those from lower mass satellites.

As for eccentricities (Fig. \ref{fig:apo-ecc}), we find no trend for $L_z$ with mass or merger redshift. However the one exception is late ($z<0.5$), massive ($M_\ast > 10^{9.5} \Msun$) mergers, which generally show prograde motion due to the spin axis of the galaxy becoming aligned with the axis of the merger.
The range in $L_z$ increases for lower mass mergers, simply because GCs from lower mass mergers have higher energies, increasing the possible range in $L_z$.

\subsection{Individual accreted galaxies}
\label{sec:indiv_gals}

\begin{figure*}
  \includegraphics[width=\textwidth]{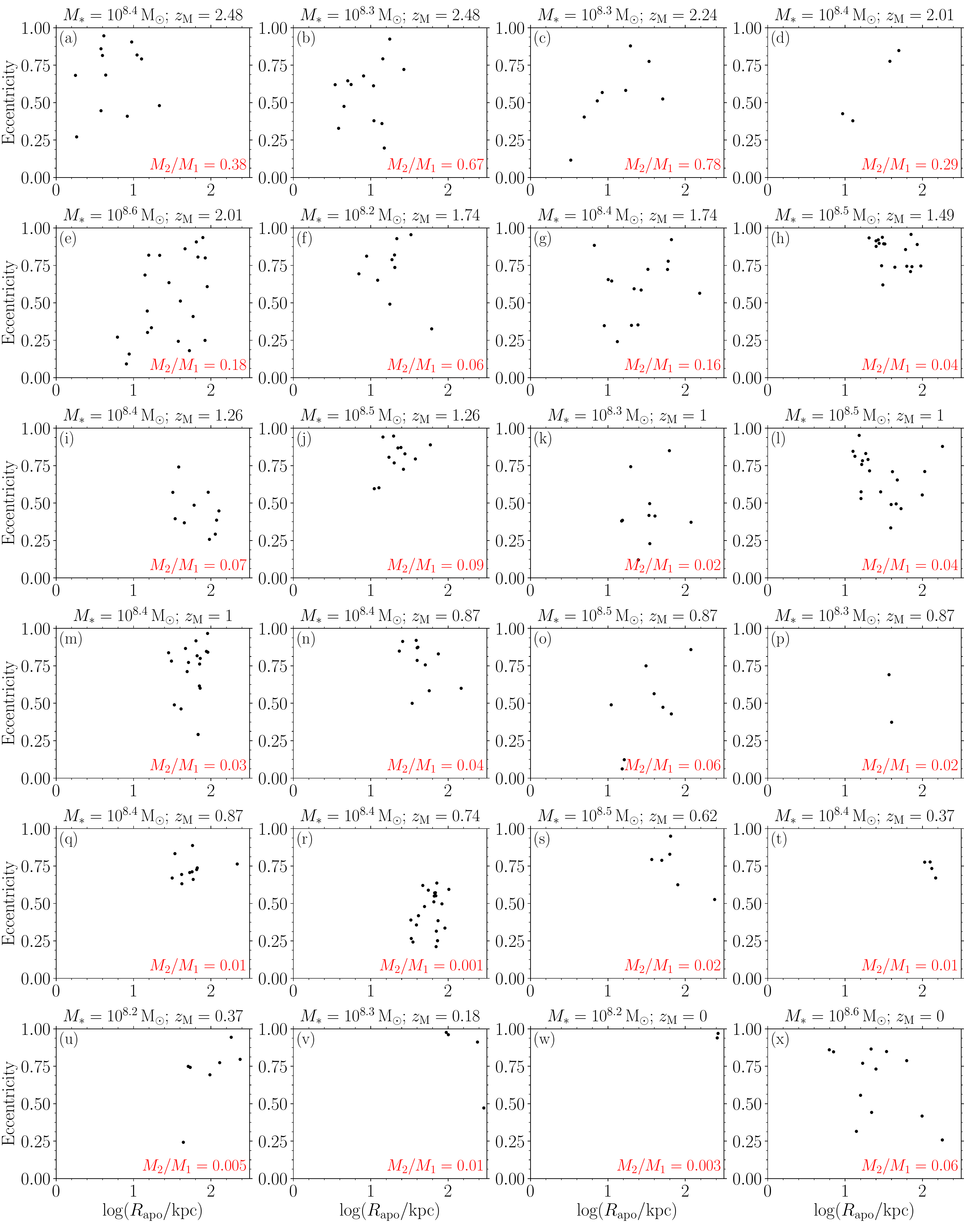}
  \caption{Eccentricity and apocentre for GCs of individual accreted galaxies with stellar masses between $10^{8.1}$ to $10^{8.6} \Msun$. Galaxies are ordered in the figure by decreasing merger redshift (left to right, top to bottom). The stellar mass merger ratio is shown in the bottom right of each panel.}
  \label{fig:gal_apo-ecc}
\end{figure*}

\begin{figure*}
  \includegraphics[width=\textwidth]{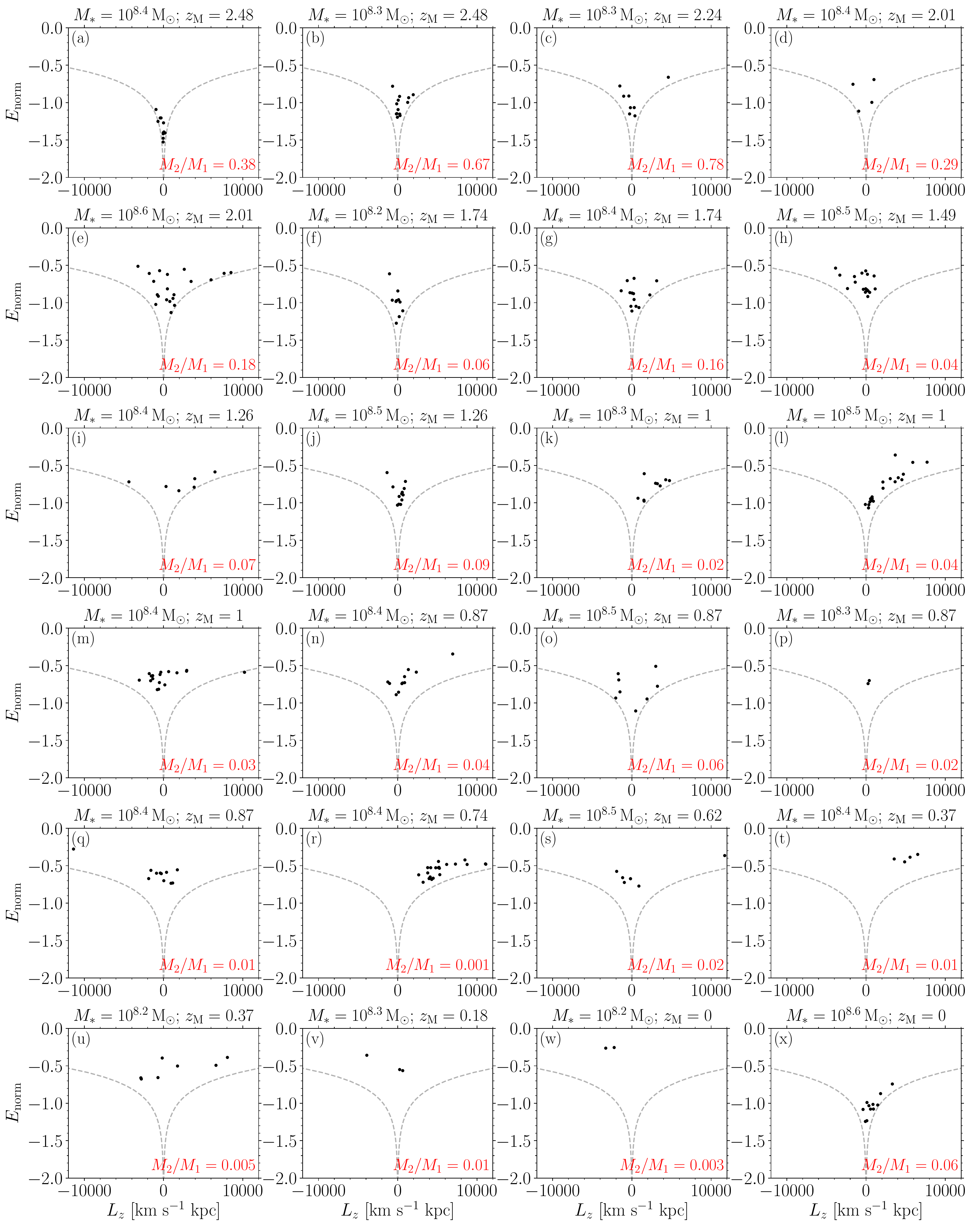}
  \caption{Normalised total energy ($E_\rmn{norm}$) and the $z$-component of the angular momentum ($L_z$) for the GCs of individual accreted galaxies as in Fig. \ref{fig:gal_apo-ecc}. The dashed lines approximately indicate the circular orbit curve for reference.}
  \label{fig:gal_E-Lz}
\end{figure*}

\begin{figure}
  \includegraphics[width=84mm]{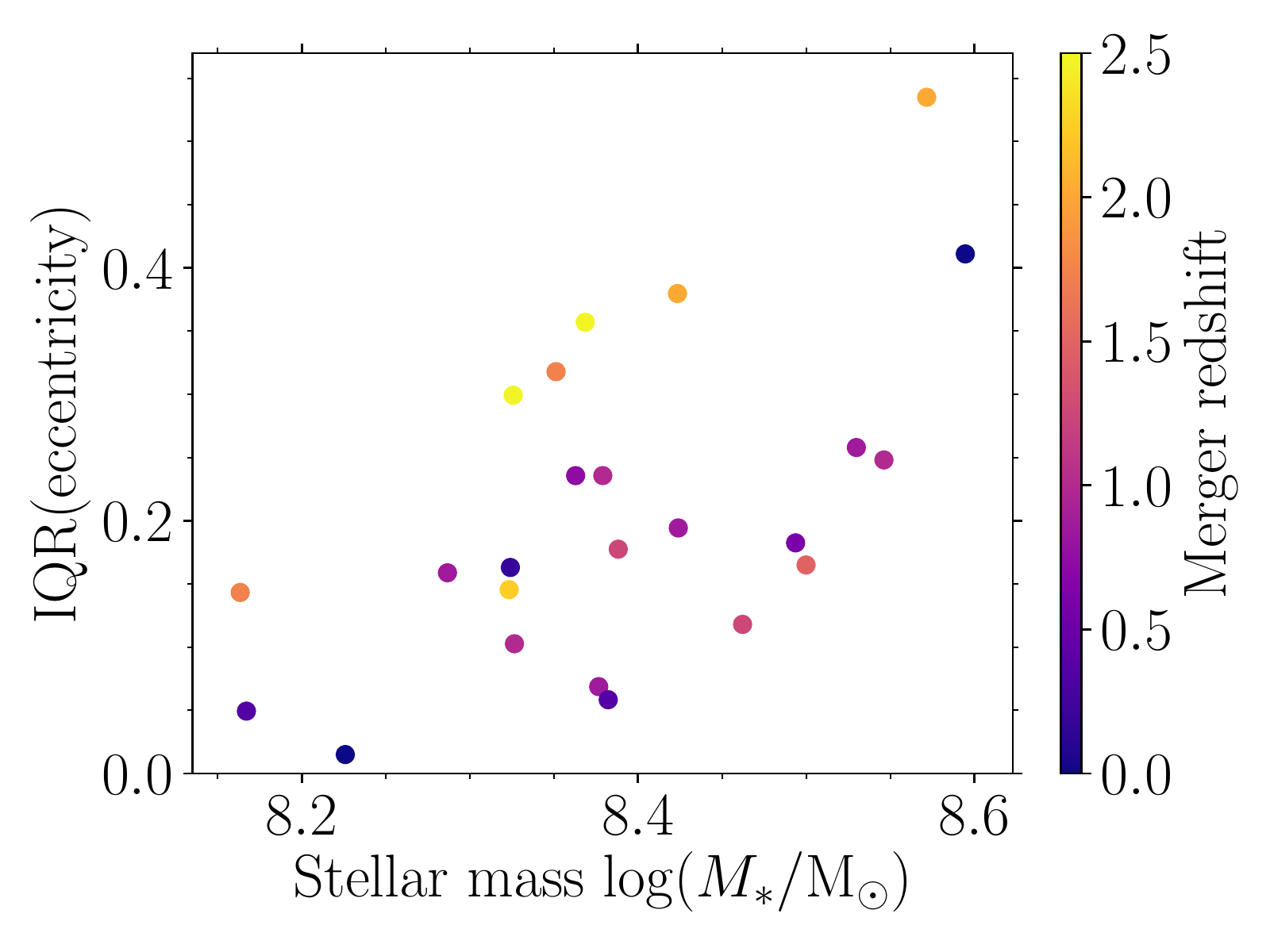}
  \caption{The interquartile range (IQR) of GC eccentricities as a function of galaxy stellar mass for the galaxies in Fig. \ref{fig:gal_apo-ecc}, with points coloured by the galaxy merger redshift. Massive and early mergers tend to have a larger range of eccentricities.}
  \label{fig:IQRecc}
\end{figure}

In Figures \ref{fig:gal_apo-ecc} and \ref{fig:gal_E-Lz} we show the apocentre-eccentricity and $E_\rmn{norm}$-$L_z$ projections, respectively, for GCs of individual accreted galaxies at $z=0$. 
The figures show accreted galaxies with stellar masses $\sim 10^{8.4} \Msun$, similar to the masses which we find for \textit{Gaia} Sausage/Enceladus and Kraken in \citet{Kruijssen_et_al_20} and that of the Sagittarius dwarf \citep{Niederste-Ostholt_et_al_10}.
Mergers often exhibit a tail of GCs to higher apocentres and energies (e.g. panels \textit{a}, \textit{e}, \textit{g} and \textit{l}). Such objects would be problematic when dividing GCs into accretion groups based on orbital properties, since any overlap with other accretion events may result in incorrect group association.

As discussed in Section \ref{sec:trends}, for a given satellite mass, earlier merger events tend to result in smaller apocentres (Fig. \ref{fig:gal_apo-ecc}) and lower total energies (Fig. \ref{fig:gal_E-Lz}).
The galaxies in Fig. \ref{fig:gal_apo-ecc} also show a diversity in the extent of their GC eccentricity distributions.
In Fig. \ref{fig:IQRecc} we compare the interquartile range (IQR) of GC eccentricities with the stellar mass of all galaxies in the figure.
Higher mass galaxies tend to have a larger IQR of GC eccentricities.
This could be the result of a broader distribution of velocities of GCs within massive galaxies, or the merger process for massive galaxies significantly altering GC orbits (the snapshot frequency of the simulations is not sufficient to investigate this further).
We also colour the points in the figure by the galaxy merger redshift. At a given galaxy mass, early mergers also tend to result in a larger range of eccentricities at $z=0$.
This could be the result of frequent mergers in the early Universe dynamically heating the orbits, increasing the spread in eccentricities, while later mergers have had significantly less time for such a process.
Alternatively (or additionally), higher merger ratios (i.e. earlier mergers at fixed galaxy mass) may simply generally result in a larger distribution of eccentricities.

With an IQR for the GC eccentricities of $0.3$ and the requirement of an early merger ($z>2$) from the apocentres of the GC orbits (Section \ref{sec:apo-ecc}), this result therefore favours an accretion event with a mass $M_\ast \gtrsim 10^{8.3} \Msun$ for the L-E group.
G-E GCs have an IQR for eccentricities of $0.18$ which suggests a mass $M_\ast \lesssim 10^{8.5} \Msun$, though it gives no constraint on accretion time.
Accretion events with tightly clustered eccentricities (IQR $\sim 0.1$; e.g. Sagittarius in Fig. \ref{fig:MW_GCs}, with an IQR of $0.07$), generally occur at later times ($z_M \lesssim 1$;) and at larger apocentres ($> 10 \kpc$). 
Of the galaxies in Fig. \ref{fig:gal_apo-ecc}, panels \textit{t} ($\mathrm{IQR} = 0.06$), \textit{v} ($\mathrm{IQR} = 0.16$) and \textit{w} ($\mathrm{IQR} = 0.02$) have accreted galaxies producing tidal streams at $z=0$ \citep[cross-matching with the list of galaxies producing streams from][and rerunning the analysis for those galaxies not previously included in their sample]{Hughes_et_al_19}.
This implies that earlier accretion events with a small distribution in eccentricity (e.g. panel \textit{h}, $\mathrm{IQR} = 0.16$) also produced tidal streams which have since dispersed.

In Fig. \ref{fig:gal_E-Lz}, GCs of a given accreted galaxy are generally tightly clustered in normalised energy.
The typical IQR in $E_\rmn{norm}$ for GCs of a given galaxy is $0.2$ (and ranges from $0.04$ to $0.34$), which does not vary with merger redshift, though some merger events exhibit a tail of GCs to higher energies (e.g. panel \textit{l}). 
This typical IQR in $E_\rmn{norm}$ is consistent with the range in energies for G-E (IQR of $0.2$) and L-E GCs (IQR of $0.09$) and further suggests they are indeed separate merger events.

\subsection{In situ GCs}
\label{sec:insitu}

\begin{figure}
  \includegraphics[width=84mm]{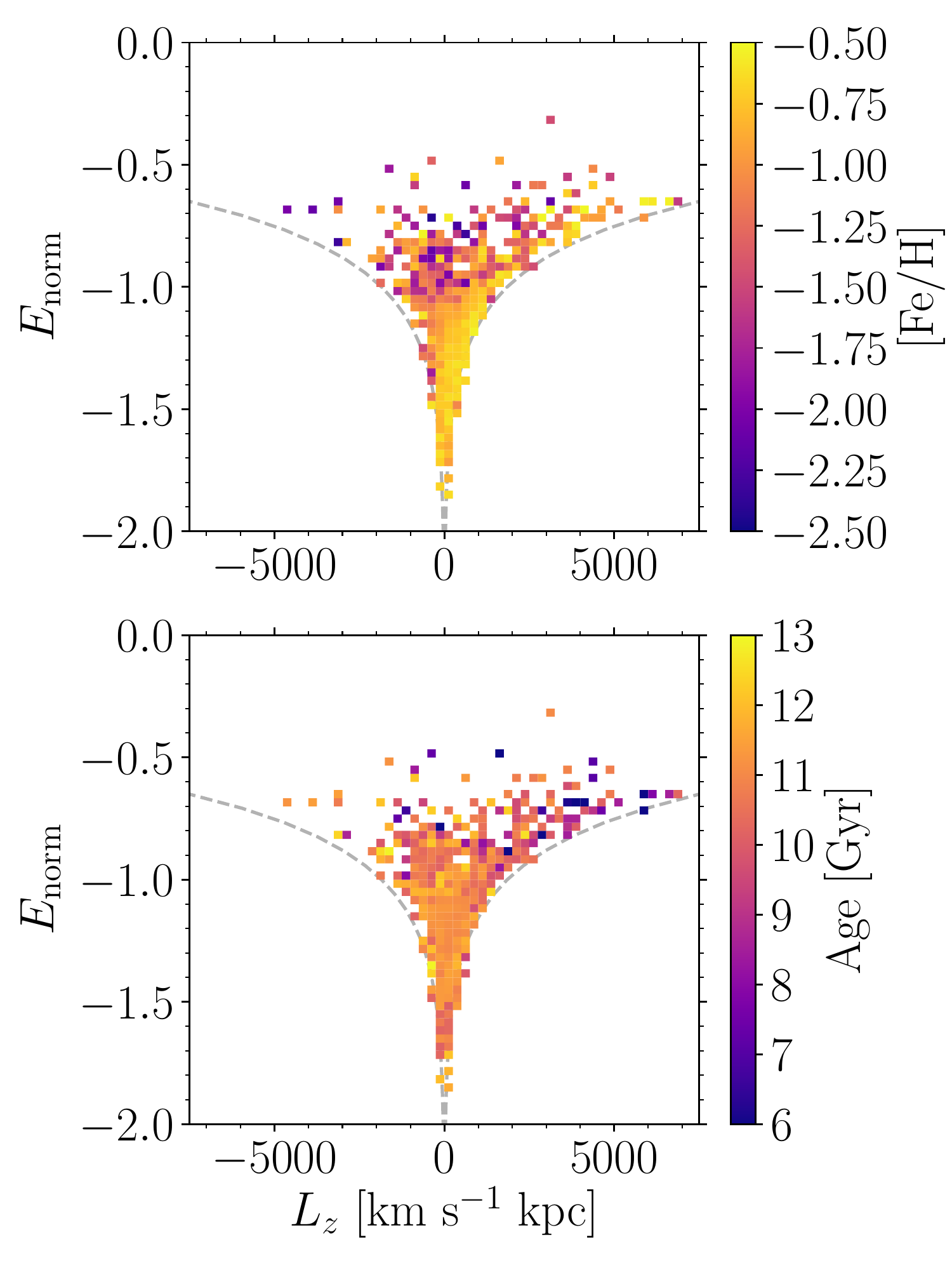}
  \caption{Normalised total energy ($E_\rmn{norm}$) and the $z$-component of the angular   momentum ($L_z$) for in situ GCs (as in the bottom left panel of Fig. \ref{fig:E-Lz}) coloured by the median metallicity (upper panel) and median age (lower panel) of each cell in the 2D histogram.}
  \label{fig:InSitu_Metals}
\end{figure}

In the bottom left panels of Figures \ref{fig:apo-ecc} and \ref{fig:E-Lz}, we compare the orbital properties of GCs formed in situ within the central galaxies in the simulations.
In order to limit the comparison to morphological analogues of the Milky Way, we only compare the GCs of galaxies which are disc-dominated ($D/T > 0.45$) at $z=0$ and have not had a major merger since $z<0.8$.

The bottom left panel of Fig. \ref{fig:apo-ecc} shows the comparison of apocentre and eccentricity.
The in situ GCs are generally very centrally concentrated, with a median apocentre of $4.3 \kpc$. 
They have a median eccentricity of $0.49$, but with a large internal spread, spanning the full eccentricity range.
The 14 galaxies individually span a range in median eccentricity of $0.42$-$0.58$.
This range is in very good agreement with the median for in situ Milky Way GCs (Fig. \ref{fig:MW_GCs}), for which the combined M-D$+$M-B sample has a median eccentricity of $0.48 \pm 0.03$.
It is unlikely this result could be affected by dynamical friction (which is not included for the orbits of the simulated GCs), since $N$-body simulations which include a live host galaxy do not find strong circularization of orbits by dynamical friction \citep{van_den_Bosch_et_al_99, Hashimoto_Funato_and_Makino_03}.

A related question is whether eccentricities of GCs are set at the time of formation, or if clusters formed on nearly circular orbits which later became more eccentric (e.g. due to galaxy mergers).
Due to the frequency of the snapshots, we cannot calculate orbits for all clusters at the time of formation. 
Limiting the sample to GCs formed $<20 \Myr$ prior to a snapshot ($N=96$ GCs, compared with 2703 for the total sample in the 14 galaxies), we find that the initial orbits (median eccentricity $0.48$) are only marginally more circular than the orbits of the GCs at $z=0$ (median $0.51$).
Thus, the median eccentricities of in situ GCs in the disc-dominated galaxies change very little over time.
However, we note that the orbits of individual GCs may significantly change between formation and $z=0$; in fact, the initial and final eccentricities of individual GCs are not correlated.
This simply follows from galactic dynamics, such that individual orbits can change substantially even if the population statistics remain the same. 

The comparison between $E_\rmn{norm}$ and $L_z$ for the in situ GCs in the disc-dominated simulations is shown in the bottom left panel of Fig. \ref{fig:E-Lz}.
In situ GCs have predominantly prograde rotation (as expected) and generally lower energies than accreted clusters \citep[cluster formation is biased to the galactic centre, where gas pressures are generally highest, see][]{P18}. 
However, in situ GCs also show a tail to high energies, and some even exhibit very retrograde orbits (counter-rotating relative to the disc).
In Fig. \ref{fig:InSitu_Metals}, we compare the median metallicity (upper panel) and age (lower panel) for in situ GCs in each cell of the $E_\rmn{norm}$-$L_z$ histogram.
The high energy and retrograde in situ GCs tend to have low metallicities ($\FeH \lesssim -1$) and formed in the early universe (ages $\sim 12 \Gyr$) when significant galaxy mergers were common, enabling the redistribution of cluster orbits or, in the case of very significant mergers, potentially changing the orientation of the angular momentum vector of the galaxy (invalidating the assumption of conservation of $L_z$).
A small number of retrograde GCs at high energies also have very young ages ($\sim 6 \Gyr$), most likely being misclassified GCs formed from gas accreted from infalling satellites.

Old, low-metallicity, in situ GCs therefore show significant overlap with accreted GCs in $E$-$L_z$ space.
In situ and accreted GCs at low metallicities also overlap in their old ages \citep{K19a} and $\alpha$-abundances \citep{Hughes_et_al_20}, meaning a combination of orbits, ages and chemistry may not be sufficient to unambiguously distinguish the origin of individual GCs in such cases \citep[see also][]{Koch_and_Cote_19}. 

In situ GCs at low energies or on prograde orbits are generally relatively metal rich ($\FeH \sim -0.5$, i.e. near the upper metallicity limit we adopt).
The GCs on nearly circular, prograde orbits at high energies and high $L_z$ also tend to be relatively young (ages $<10 \Gyr$) compared to the GCs at low energies (ages $\sim12 \Gyr$), due to the inside-out nature of disc formation \citep[e.g.][]{Larson_76, Matteucci_and_Francois_89, Burkert_Truran_and_Hensler_92, Munoz-Mateos_et_al_07}.
For in situ GCs in the simulations, outer disc GCs with apocentres $>10 \kpc$ (at $z=0$) represent $\approx 11$ per cent of GCs with very circular orbits (eccentricities $<0.3$).
These objects have a number of possible analogues in the Milky Way, namely Palomar 1, Palomar 5, E3 and NGC 5053 \citep{Baumgardt_et_al_19}. The four GCs make up 18 per cent of all Milky Way GCs with eccentricities $<0.3$.
In the \citet{Massari_et_al_19} list of possible progenitor galaxy associations, E3 was associated with the main disc, NGC 5053 and Palomar 5 are possible associations with the H99 streams and Palomar 1 was listed in the `high energy' group (no known progenitor).
Of the four, Palomar 1 could be an example of a GC formed in situ in the outer Milky Way disc, given its age of $\sim 7 \Gyr$ and metallicity $\FeH = -0.7$ \citep{Forbes_and_Bridges_10}.
Palomar 1 has also previously been classified as a bulge/disc cluster based on its horizontal branch morphology \citep{Zinn_93, Mackey_and_van_den_Bergh_05}.

\section{Discussion and conclusions}
\label{sec:discussion}

Analysing the results of the E-MOSAICS simulations of Milky Way-mass galaxies, we find that the orbits (apocentre and total energy, in particular) of GCs deposited by accretion events are sensitive to the satellite galaxy mass and merger redshift. 
Earlier mergers and larger galaxy masses result in more tightly bound GCs (smaller apocentres).
We expect these trends should exist across all host galaxy masses, though the exact relationships between the apocentre or energy of the orbits and satellite mass and merger redshift will differ with host galaxy/halo mass.

Taking advantage of the GC groupings corresponding to probable accretion events defined by \citet{Massari_et_al_19}, we estimate merger redshifts based on the apocentres of the GC orbits and the most likely progenitor stellar masses:

\begin{itemize}

\item For \textit{Gaia} Sausage/Enceladus we find a merger redshift in the range $z_M \approx 1$-$2$, depending on the assumed stellar mass for the progenitor ($10^{7.5} < M_\ast/\Msun < 10^{9.5}$).
The small interquartile range of eccentricities for the G-E group of GCs favour an accretion event with stellar mass $M_\ast \lesssim 10^{8.5} \Msun$ (Section \ref{sec:indiv_gals}).
This is in reasonable agreement with \citet{Belokurov_et_al_18}, who suggest a merger 8-11 Gyr ago based on the velocity anisotropy of the stellar debris, and \citet{Helmi_et_al_18}, who suggest a merger $\approx10 \Gyr$ ago based on the youngest G-E stars.
It is also consistent with the results of \citet{Mackereth_et_al_19}, who found that galaxy accretion resulting in very eccentric orbits (median eccentricity for the stellar debris of $> 0.8$) only occur for accretion at late times, implying an accretion redshift of $z<1.5$.
We note that \citet{Bignone_Helmi_and_Tissera_19} found a possible G-E analogue in the EAGLE simulations, for which the merger occurred at $z \sim 1.2$.

\item For the H99 stream \citep{Helmi_et_al_99} and Sequoia \citep{Myeong_et_al_19} we find $z_M \sim 2$ and $1.5$, respectively (and certainly $z_M > 1$).
The implied merger times are consistent with the ages of the youngest GCs in each group  and the youngest stars ($\approx 11 \Gyr$ old) in H99 \citep{Koppelman_et_al_19a, Massari_et_al_19, Myeong_et_al_19}. 
Based on idealised $N$-body simulations, \citet{Kepley_et_al_07} and \citet{Koppelman_et_al_19a} suggest the H99 stream progenitor was accreted 5-9 Gyr ago (with lower galaxy masses implying older mergers).
Assuming Sequoia and the S1 stream are connected, \citet{Myeong_et_al_18a, Myeong_et_al_19} suggest an infall time for Sequoia of $>9 \Gyr$ ago, in agreement with our result.

\item The median apocentre of Sagittarius GCs implies a late merger ($z_M < 1$), in agreement with the dwarf galaxy currently undergoing tidal disruption \citep{Ibata_Gilmore_and_Irwin_94}.

\item The L-E group \citep{Massari_et_al_19}, for which the progenitor stellar debris is yet to be discovered, has the most compact apocentres (median $4.4 \kpc$) of all (presumably) accreted subpopulations (Fig. \ref{fig:MW_GCs}).
Such small apocentres require (Table \ref{tab:apo}) an early accretion time ($z_M \gtrsim 2$) and a progenitor galaxy mass $>10^{7.5} \Msun$.
Combined with the constraint on merger redshift, the range of GC eccentricities in the L-E group favour an accretion event with stellar mass $M_\ast \gtrsim 10^{8.3} \Msun$ (Section \ref{sec:indiv_gals}).
This is in agreement with the age-metallicity relation of the L-E group GCs \citep{Massari_et_al_19} which indicates a relatively massive progenitor galaxy \citep{K19a,K19b} and disfavours in situ formation in the Galaxy due to their low metallicities at ages $\approx 11 \Gyr$.
The implied merger time is also consistent with the age ($\approx 10.5 \Gyr$) of the youngest L-E GC \citep[or $\approx 11 \Gyr$ for the second youngest GC]{Massari_et_al_19}.

\end{itemize}

These results reaffirm the findings of \citet{K19b} that the Milky Way underwent two massive accretion events in its past.
We argue that the L-E group is in fact remnants of the Kraken event \citep[see also][]{Forbes_20,Kruijssen_et_al_20}, predicted by \citet{K19b} based on the number of GCs in the satellite branch of the GC age-metallicity distribution and the galaxy mass implied by their age-metallicity relation.
The large distribution of eccentricities and small apocentres of Kraken GCs (and thus presumably also its field stars) means that detecting the stellar debris of the merger may be a tough prospect, since there will not be an obvious clustering of stars in orbital space and there may be significant overlap with Milky Way disc stars. 
Alternatively, with accurate stellar ages, it may be possible to find stars on tightly bound orbits with properties (ages $\approx 11 \Gyr$, $\FeH \approx -1.3$) similar to the youngest Kraken GCs.
Such stars should be significantly younger than in situ stars at similar metallicities (or more metal-poor than in situ stars at similar ages).

Finally, we investigate the orbits of in situ GCs in the simulated galaxies.
We find that the median eccentricities of both in situ ($0.49$) and accreted GCs ($0.71$) in the simulations are in remarkable agreement with the Milky Way GCs ($0.48$ for M-D$+$M-B GCs and $0.7$ for all other sub-groups combined).
This result provides further evidence that a formation mechanism similar to that observed for young star clusters \citep[for reviews, see][]{Portegies-Zwart_McMillan_and_Gieles_10, Kruijssen_14, Adamo_and_Bastian_18}, combined with hierarchical formation and assembly of galaxies, can explain the GC populations observed today \citep[e.g.][]{Elmegreen_and_Efremov_97, Kravtsov_and_Gnedin_05, Kruijssen_15, Li_et_al_17, P18, K19a, Lahen_et_al_19, Ma_et_al_20}.

We also find there can be significant overlaps in orbital properties between in situ and accreted GCs.
Though the in situ GCs are generally biased to low energies (small apocentres), they exhibit a large tail to high energies and even retrograde orbits (relative to the present-day disc), such that there is a significant overlap between in situ and accreted GCs.
The high-energy in situ GCs are generally old and metal-poor, meaning it may not be possible to unambiguously distinguish between in situ and accreted GCs in these cases.

We find that the orbits of GC subpopulations may hold particular power in recovering the properties of their progenitor galaxy, though there exists a degeneracy between the galaxy mass and the accretion redshift when considering only orbital properties.
In this paper we rely on existing estimates for galaxy masses to derive merger redshifts, breaking this degeneracy therefore requires combining the orbital properties with other tracers.
We undertake this in a companion paper \citep{Kruijssen_et_al_20}, combining the GC subpopulation orbits with their ages and metallicities to recover their progenitor galaxy properties using information about the GC sub-systems alone.
All of our estimated accretion redshifts in this paper are consistent to within the formal uncertainties with the predictions of \citet{Kruijssen_et_al_20}, however the extra age and metallicity information used in that paper allows us to simultaneously derive progenitor galaxy masses.
The combination of age-metallicity information with orbital properties therefore greatly increases the power of GCs in their use as tracers of galaxy formation and assembly.

\section*{Acknowledgements}

We thank the referee for a helpful and constructive report.
JP and NB gratefully acknowledge funding from a European Research Council consolidator grant (ERC-CoG-646928-Multi-Pop). 
JMDK gratefully acknowledges funding from the Deutsche Forschungsgemeinschaft (DFG, German Research Foundation) through an Emmy Noether Research Group (grant number KR4801/1-1). JMDK, STG, and MRC gratefully acknowledge funding from the European Research Council (ERC) under the European Union's Horizon 2020 research and innovation programme via the ERC Starting Grant MUSTANG (grant agreement number 714907). MRC is supported by a Fellowship from the International Max Planck Research School for Astronomy and Cosmic Physics at the University of Heidelberg (IMPRS-HD).
NB and RAC are Royal Society University Research Fellows.
This work used the DiRAC Data Centric system at Durham University, operated by the Institute for Computational Cosmology on behalf of the STFC DiRAC HPC Facility (\url{www.dirac.ac.uk}). This equipment was funded by BIS National E-infrastructure capital grant ST/K00042X/1, STFC capital grants ST/H008519/1 and ST/K00087X/1, STFC DiRAC Operations grant ST/K003267/1 and Durham University. DiRAC is part of the National E-Infrastructure.
The work also made use of high performance computing facilities at Liverpool John Moores University, partly funded by the Royal Society and LJMU's Faculty of Engineering and Technology.

\section*{Data Availability}

The data underlying this article will be shared on reasonable request to the corresponding author.



\bibliographystyle{mnras}
\bibliography{emosaics}






\bsp	
\label{lastpage}
\end{document}